\documentclass[aps,twocolumn,prl,superscriptaddress,showpacs,showkeys,floatfix]{revtex4-1}

\usepackage{amsbsy}
\usepackage{graphicx}
\usepackage{amsmath}
\usepackage{amssymb}
\usepackage{bm}
\input{epsf}

\begin{document}

\title{Dynamical Casimir effect and minimal temperature in quantum thermodynamics}

\author{Giuliano Benenti}
\affiliation{Center for Nonlinear and Complex Systems,
Universit\`a degli Studi dell'Insubria, via Valleggio 11, 22100 Como, Italy}
\affiliation{Istituto Nazionale di Fisica Nucleare, Sezione di Milano,
via Celoria 16, 20133 Milano, Italy}
\author{Giuliano Strini}
\affiliation{Department of Physics, University of Milan,
via Celoria 16, 20133 Milano, Italy}

\begin{abstract}
We study the fundamental limitations of cooling to absolute zero for
a qubit, interacting with a single mode of the electromagnetic field. Our
results show that the dynamical Casimir effect, which is unavoidable in 
any finite-time thermodynamic cycle, forbids the attainability of the 
absolute zero of temperature, even in the limit of an infinite
number of cycles.
\end{abstract}


\maketitle


\textit{Introduction.}
Due to recent progress
of nanofabrication technology, quantum effects in small
heat engines have become an increasingly important subject.
Concepts from quantum thermodynamics~\cite{mahlerbook} 
have been applied to investigate questions 
such as the optimization of quantum thermal machines~\cite{benenti2013},
the fundamental dimensional limits to thermodynamic machines~\cite{popescu},
and the minimum temperature achievable in nanoscopic 
chillers~\cite{kosloffEPL2009,kosloffEPL2010,eisert2012,vandenbroeck2012,kurizki2012,kosloff2012,kosloff2012b,kosloff2013}.  
Cooling a system to the absolute zero of temperature ($T=0$) 
is prohibited by Nernst's unattainability principle~\cite{nerst}, 
also known as the dynamical formulation of the third law of thermodynamics.
Such principle states that it is impossible by any 
procedure to reduce any system to $T=0$ in finite time.
Nernst's principle has been recently challenged~\cite{kurizki2012}. 

It this Letter, we investigate the unattainability principle
in a minimal model: a qubit coupled to a single mode of
the electromagnetic field, i.e. a harmonic oscillator. 
The oscillator is the working medium, shuttling heat from the
qubit to a hot reservoir by means of a (quantum) Otto cycle. 
Although oversimplified, our model has several relevant features:
\begin{itemize}
\item
The matter-field interaction is treated at the fundamental level
of quantum electrodynamics.
\item
The equations of motion are solved by accurate numerical simulations,
without resorting to the approximations necessarily involved 
in the master equations often used in the literature;
\item
We do not use the Rotating Wave Approximations (RWA), that
neglects the effects of rapidly rotating terms in the equations of
motion~\cite{micromaser}. 
\end{itemize}
The latter point is particularly relevant where addressing the 
fundamental limits to cooling, since the terms beyond the RWA lead
to the generation of photons from the vacuum due to time-dependent
boundary conditions for the electromagnetic field. Such 
\emph{quantum} vacuum amplification effect,
known as the Dynamical Casimir Effect (DCE)~\cite{moore,dodonov,noriRMP},
has been observed in recent experiments with superconducting 
circuits~\cite{norinature,lahteenmaki}, 
and also investigated in the context of
Bose-Einstein condensates~\cite{jaskula}, in excition-polariton
condensates~\cite{koghee}, for multipartite entanglement generation
in cavity networks~\cite{solano2014} and for quantum  
communication protocols~\cite{casimirqip}.
Here, we show that the DCE is 
a fundamental, purely quantum, limitation to cooling.
We point out that the DCE is unavoidable in the context of finite-time
thermodynamics~\cite{andresen} for cyclic quantum cooling machines, where  
the field's boundary conditions are effectively changed in time
due to the switching on/off of the matter-field coupling~\cite{footnote}. 
In what follows, we show that, due to the DCE, the T=0 state
of the qubit cannot be attained, even in the limit of infinite 
number of cooling cycles. 


\textit{The model.}
We consider a reciprocating refrigerator, operating by means of a 
working medium [a single mode of the electromagnetic field, that is,
a harmonic oscillator, with a time-dependent frequency $\omega(t)$], 
shuttling heat from a cold finite-size ``bath'' (a single qubit) to a hot bath.
The working medium undergoes a four-stroke Otto cycle:
\begin{itemize}
\item 
\emph{Isochore} A$\to$B: the working medium is in contact with the cold 
bath at temperature $T_c$; the work parameter, i.e. 
the oscillator frequency is maintained constant,
$\omega(t)=\omega_c$.
\item
\emph{Adiabatic compression} B$\to$C: the frequency $\omega(t)$ of 
the working medium changes in time from $\omega_c$ to $\omega_h$.
\item
\emph{Isochore} C$\to$ D: the working medium is in contact with the 
hot bath at temperature $T_h$; the oscillator frequency is constant,
$\omega(t)=\omega_h$.
\item
\emph{Adiabatic expansion} D$\to$A: the frequency $\omega(t)$ changes 
from $\omega_h$ to $\omega_c$.
\end{itemize}

In order to elucidate the limitations imposed by the 
DCE on the lowest temperature that 
could be reached by a cooling protocol, we first consider 
the isochore stroke A$\to$B, with both the qubit and the
oscillator prepared in their ground state.   
We wish to emphasize here that, even starting from these
ideal conditions, due to the DCE 
both the oscillator and the qubit are excited, so that 
at the end of the isochore stroke the qubit is left in 
a state at a nonzero temperature.

During the stroke A$\to$B
the qubit-oscillator interaction is described by the
time-dependent Rabi Hamiltonian~\cite{micromaser} (we set 
the reduced Planck's constant $\hbar=1$):
\begin{equation}
  \begin{array}{c}
{\displaystyle
H(t)=H_0+H_I(t),
}
\\
\\
{\displaystyle
H_0=-\frac{1}{2}\,\omega_a \sigma_z +
\omega\left(a^\dagger a +\frac{1}{2}\right),
}
\\
\\
{\displaystyle
H_I(t)=f(t)\,\left[g \,\sigma_+\,(a^\dagger+a)
+g^\star \sigma_-\,(a^\dagger+a)\right],
}
\end{array}
\label{eq:noREWAquantum}
\end{equation}
where $\sigma_i$ ($i=x,y,z$) are the Pauli matrices,
$\sigma_\pm = \frac{1}{2}\,(\sigma_x\mp i \sigma_y)$
are the rising and lowering operators for the qubit:
$\sigma_+ |g\rangle = |e\rangle$,
$\sigma_+ |e\rangle = 0$,
$\sigma_- |g\rangle = 0$,
$\sigma_- |e\rangle = |g\rangle$;
the operators $a^\dagger$ and $a$ for the field create
and annihilate a photon:
$a^\dagger |n\rangle=\sqrt{n+1}|n+1\rangle$,
$a |n\rangle=\sqrt{n}|n-1\rangle$,
$|n\rangle$ being the Fock state with $n$ photons.
We first assume sudden switch on/off of the coupling:
$f(t)=1$ for $0\le t \le \tau$, $f(t)=0$ otherwise,
with $\tau$ duration of the A$\to$B stroke.
For simplicity's sake, we
consider the resonant case ($\omega=\omega_a$) and
the coupling strength $g\in\mathbb{R}$.
The RWA (exact only in the limit $g\to0$)
is obtained when we neglect the term
$\sigma_+ a^\dagger$, which simultaneously
excites the qubit and creates a photon,
and $\sigma_- a$, which de-excites the
qubit and annihilates a photon. In this limit, Hamiltonian
(\ref{eq:noREWAquantum}) reduces to the Jaynes-Cummings
Hamiltonian \cite{micromaser}. 
We set $\omega=1$, so that in the RWA the swap time needed to transfer
an excitation from the qubit to the field or vice versa
($|e\rangle |0\rangle\leftrightarrow |g\rangle |1\rangle$)
is $\tau_S=\pi/2g$.

\begin{figure}[htp]
\includegraphics[angle=0.0, width=8cm]{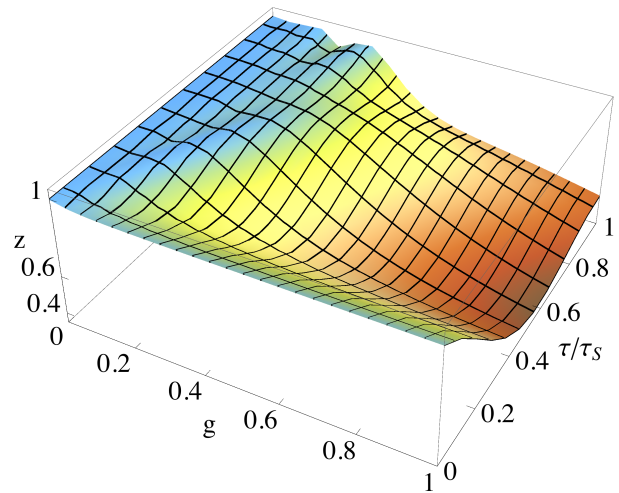}
\caption{(color online) Bloch coordinate $z$ of the qubit after the A$\to$B isochore,
as a function of the qubit-oscillator interaction time $\tau$ 
(in units of the swap time $\tau_S$) and
the interaction strength $g$.}
\label{fig:gg}
\end{figure}

\textit{Results.}
If the qubit is prepared in its ground state, $\rho_q(0)=|g\rangle\langle g|$ 
and the oscillator is in the vacuum state, $\rho_o(0)=|0\rangle\langle 0|$,
then, as we will discuss below, the qubit remains in a diagonal state
in the basis of its eigenstates,
$\rho_q(t)= p(t) |g\rangle\langle g|+ [1-p(t)] |e\rangle\langle e|$.
Therefore, the qubit's temperature reads as follows:    
$T(t)=\omega/\ln \{p(t)/[1-p(t)]\}$ (we set the Boltzmann constant $k_B=1$).
We found more convenient for visualization to plot, rather than the
temperature, the $z$-coordinate of the Bloch vector~\cite{qcbook} 
${\bf r}(t)=[x(t),y(t),z(t)]$ of the 
state $\rho_{q}(t)=\frac{1}{2}\left(I+{\bf r}(t)\cdot \bm{\sigma}\right)$, 
with $I$ identity operator and $\bm{\sigma}=(\sigma_x,\sigma_y,\sigma_z)$.
In Fig.~\ref{fig:gg}, we show $z(\tau)$ at the end of the isochore transformation 
A$\to$B, as a 
function of $\tau$ and of the qubit-oscillator coupling strength $g$.
Within the RWA the initial tensor-product state 
$\rho_q(0)\otimes \rho_o(0)$ is the ground state of the overall
qubit-oscillator system. Hence, $z(\tau)=1$, that is, the temperature $T(\tau)=0$ 
for any value of $\tau$.
On the other hand, due to the DCE for any value of $g$ there exists a finite 
probability to generate photons and to excite the qubit, so that
$z(\tau)<1$, namely the temperature $T(\tau)$ of the qubit is nonzero. 

%
%

Since we are interested in the fundamental limitations to
cooling imposed by the DCE, 
for the remaining part of the Otto cycle we ideally consider 
the most favorable conditions for cooling.
That is, we assume that the adiabatic transformations
can be performed without friction in a finite time 
by utilizing ``shortcuts to adiabaticity''~\cite{torrontegui,paternostro}.
Furthermore, we assume that the overall cycle is engineered in
such a way that the oscillator is left in its vacuum state 
at the end of the cycle. 
While a careful treatment of the Otto cycle should be performed
to evaluate the cooling power of a 
refrigerator~\cite{kosloffEPL2009,kosloffEPL2010}, our analysis,
based on the most favorable instance, is sufficient to 
investigate the limitations set by the DCE to the lowest 
attainable temperature in finite-time thermodynamic cycles.

\begin{figure}[htp]
\includegraphics[angle=0.0, width=8cm]{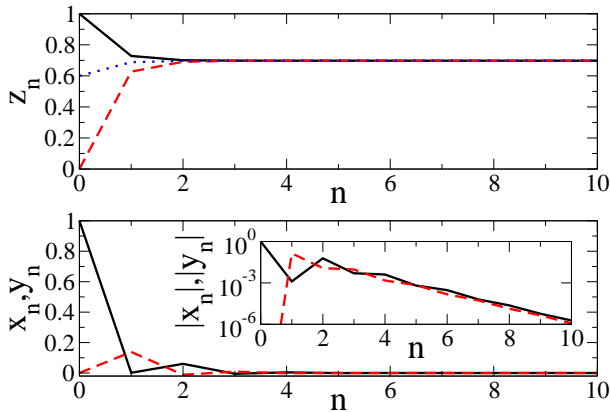}
\caption{(color online) 
Evolution of the Bloch ball coordinates as a function of 
the number of Otto cycles. 
Top: $z$-coordinate for 
the initial state of the qubit $|g\rangle\langle g|$ (black full curve),
$\frac{1}{2}\,I$ (red dashed curve),
$p|g\rangle\langle g|+
(1-p)|e\rangle\langle e|$, with $p=0.6$ (blue dotted curve).
In the bottom panel we show, for the state
$\frac{1}{\sqrt{2}}\,(|g\rangle+|e\rangle)$
the $x$- (black full curve)
and the $y$-coordinate (red dashed curve). Note that for this state
$z$ evolves exactly as for the maximally mixed input state.
In the inset, for the same initial state, we show the exponential decay of
$|x|$ and $|y|$.}
\label{fig:bloch}
\end{figure}

With the above assumptions, the state of the qubit after 
$n$ Otto cycles is given by
\begin{equation}
\rho_{q,n}={\rm Tr}_o\,[U(\rho_{q,n-1}\otimes |0\rangle\langle 0)|U^\dagger],
\label{eq:rhoqn}
\end{equation}
with $U$ the unitary time evolution operator describing the 
A$\to$B evolution for the qubit and the oscillator
and $\rho_{q,0}\equiv \rho_q(0)$ initial state of the qubit.
The quantum channel $\mathcal{E}$ mapping $\rho_{q,n-1}$
into $\rho_{q,n}$ can be conveniently described in the Fano-Bloch
representation \cite{fano,eberly,mahler,striniqpt,strinidiamondnorm,casimirqip}.
If ${\bf r}_n=(x_n,y_n,z_n)$ denotes the Bloch vector of
the state $\rho_{q,n}$,
it follows from the linearity of quantum mechanics that ${\bf r}_{n-1}$ 
and ${\bf r}_{n}$
are connected through an affine map $\mathcal{M}$:
\begin{equation}
\left[
\begin{array}{c}
{\bf r}_{n}
\\
\hline
1
\end{array}
\right]
=
\mathcal{M}
\left[
\begin{array}{c}
{\bf r}_{n-1}
\\
\hline
1
\end{array}
\right]
=
\left[
\begin{array}{ccc}
  {\bf M} & \Big \lvert & {\bf a}  \\
 \hline
  {\bf 0}^T & \Big \lvert & 1
\end{array}
\right]
\left[
\begin{array}{c}
{\bf r}_{n-1}
\\
\hline
1
\end{array}
\right],
\label{eq:affine}
\end{equation}
where ${\bf M}$ is a $3\times 3$ real matrix,
${\bf r}_{n-1}$, ${\bf r}_{n}$ and ${\bf a}$ real column vectors of
dimension $3$ and
${\bf 0}$ the null vector of the same dimension.
The Fano-Bloch representation 
of quantum operations is also physically transparent
since the Bloch vector directly provides the expectation values
of polarization measurements.
While in general an affine map for a qubit depends
on twelve parameter~\cite{qcbook},
we found from the numerical simulation of the quantum map 
$\mathcal{E}$ the following structure
of ${\bf M}$ and ${\bf a}$:
\begin{equation}
{\bf M} = \left(
\begin{array}{ccc}
  m_{xx} & m_{xy} & 0        \\
  m_{yx} & m_{yy} & 0        \\
    0    &   0    & m_{zz}   \\
  \end{array} \right), \qquad
{\bf a} = \left(
\begin{array}{c}
    0   \\
    0   \\
    a_z \\
  \end{array} \right).
\label{eq:Memoryless-Kraus-Operators}
\end{equation}
It follows that, whatever the initial state $\rho_{q,0}$ is,
for $n\to\infty$ we have $x_n,y_n\to 0$ and 
$z_n\to a_z/(1-m_{zz})$. Therefore the asymptotic state of the 
qubit, $\rho_{\infty}=\frac{1}{2}\,(1+z_{\infty})|g\rangle\langle g|+
\frac{1}{2}\,(1-z_{\infty})|e\rangle\langle e|$, is diagonal
and we can readily derive the asymptotic temperature $T_\infty$. 
In the particular case in which the state is diagonal from the 
beginning, then, as observed above, it remains diagonal and
the temperature $T_n$ can be computed as a function of the 
number $n$ of Otto cycles.
Examples of the evolution of the Bloch ball coordinates 
$x_n,y_n$ and $z_n$ are shown in Fig.~\ref{fig:bloch},
for different initial states: the qubit's ground state
$|g\rangle\langle g|$, the maximally mixed state
$\frac{1}{2}\,I$, a thermal state $p|g\rangle\langle g|+
(1-p)|e\rangle\langle e|$, with $p=0.6$, and 
the superposition state $\frac{1}{\sqrt{2}}\,(|g\rangle+|e\rangle)$.
As expected, the $z$-coordinate converges to a limiting value
depending only on the channel's control parameters 
(here, $g=0.5$ and $\tau=\pi/2g$) but not on the initial state
of the qubit. For the superposition state we also show the 
asymptotic decay of $x$ and $y$ (in the other cases $x=y=0$).    

%
%

\begin{figure}[htp]
\includegraphics[angle=0.0, width=8cm]{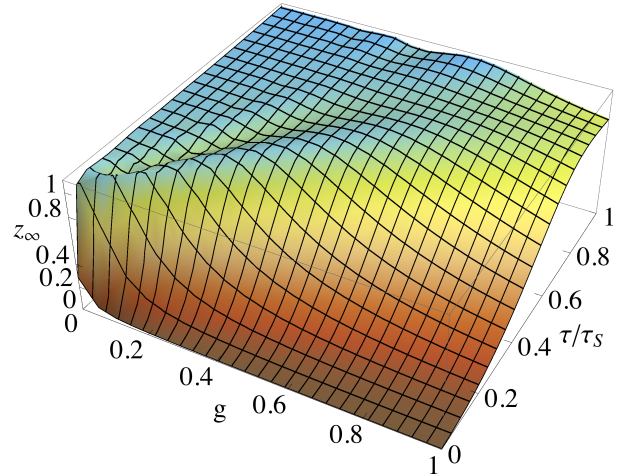}
\caption{(color online) Asymptotic value $z_\infty$ of the  
$z$-coordinate for the qubit,
as a function of the time $\tau$ and
the interaction strength $g$.}
\label{fig:Tinfty}
\end{figure}

In Fig.~\ref{fig:Tinfty} we show the asymptotic value 
$z_\infty=\lim_{n\to\infty} z_n$ 
as a function of the time $\tau$ and
of the interaction strength $g$.
It is clear from this plot that in our model 
only in the limit $g\to 0$ the temperature asymptotically vanishes.
In the same figure, it might appear at first sight surprising that
for a given $g$ and $\tau\to 0$ we have $z_\infty\to 0^+$, namely 
$T_\infty\to +\infty$. In this limit, we have 
$a_z\to 0^+$ and 
$m_{zz}\to 1^-$, while the ratio 
$z_\infty=a_z/(1-m_{zz})\to 0^+$~\cite{footnote2}. 
Note that from the affine map
(\ref{eq:affine}) we have $z_{n}=m_{zz} z_{n-1}+ a_z$, and therefore
a value of $m_{zz}$ close to $1$ implies a slow convergence to the asymptotic
temperature.

%
%

\begin{figure}[htp]
\includegraphics[angle=0.0, width=8cm]{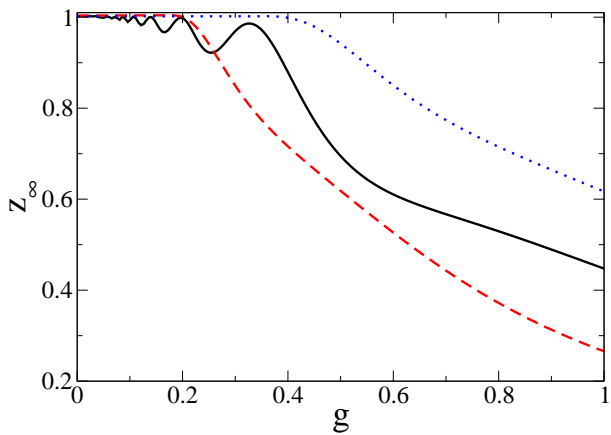}
\caption{(color online) Asymptotic value $z_\infty$ of the $z$-coordinate
for the qubit,
as a function of the interaction strength $g$,
for the rectangular (black full curve) and the
Hamming window for $\alpha=1$ (red dashed curve) or 
$\alpha=2$ (blue dotted curve).
The interaction time $\tau=\tau_S=\pi/2g$.}
\label{fig:window}
\end{figure}

Finally, we investigate the effects of a smooth switch on/off
of the interaction, by substituting in 
Eq.~(\ref{eq:noREWAquantum}) the rectangular window so far
considered with the 
Hamming window:
$f(t)=\frac{1}{\alpha}\,[1-\cos(2\pi t/\alpha\tau)]$ if $0\le t\le \alpha\tau$,
$f(t)=0$ otherwise. In particular, we consider the values $\alpha=1$, 
for which the interaction time (equal to $\tau$) is the same as 
for the rectangular window and 
the peak value (equal to $2g$) is doubled, and $\alpha=2$, for which the 
interaction time ($2 \tau$) is doubled and the peak value ($g$) unchanged.
In both cases, the area below the Hamming window is the 
same as for the rectangular window previously considered,
so that such a window does not affect the cooling within 
the RWA, while significant differences appear in the
ultra-strong coupling regime. 
In Fig.~\ref{fig:window} we consider the time $\tau=\tau_S=\pi/2g$,
so that within the RWA the asymptotic temperature $T_\infty=0$
[in the affine map~(\ref{eq:affine}), $m_{zz}=0$ and $a_z=1$]. 
More precisely, the qubit achieves the zero-temperature state
in a single Otto cycle, since for this value of $\tau$ and
for the oscillator prepared in the vacuum state, the 
qubit and the oscillator swap their state, leaving the qubit
in its ground state. 
On the other hand, if the terms beyond the RWA are taken into account, 
the asymptotic temperature changes with respect to the rectangular 
window. The oscillations of $T_\infty$ with $g$ are smoothed;
the temperature is raised for $\alpha=1$ and lowered for $\alpha=2$.  
This result can be understood from the adiabatic theorem: for $\alpha=2$
the interaction is switched on/off more gradually than for $\alpha=1$
and therefore, if we start from an eigenstate of the unperturbed Hamiltonian
$H_0$, for $\alpha=2$ we end up, after the $A\to B$ isochore, on the same 
eigenstate with higher probability than for the case $\alpha=1$.
In particular, if we start from the $T=0$ state, for the Hamming window 
such state is exactly preserved only in the limit $\alpha\to\infty$,
that is, the qubit-field interaction time $\tau_I=\alpha\tau\to\infty$.
In this limit the DCE vanishes since it is an effect due to the 
change of the boundary conditions for the field when it interacts with
the qubit and such change becomes infinitely slow. 

We should like to 
stress that the two limits of interaction time $\tau_I\to\infty$ 
and number of cycles $n\to\infty$ do not commute. If we let before 
$\tau_I\to\infty$, then we can achieve (in infinite time) the $T=0$ limit,
as allowed from Nernst's principle.
If on the other hand we set a large as desired but finite value of 
$\tau_I$ and then let $n\to\infty$, then the $T=0$ limit is not
attained. This latter case is the one relevant for cyclic quantum 
chillers in finite-time thermodynamics. 

%
%


\textit{Discussion and conclusions.}
In this paper, we have investigated the limitations
imposed by the DCE on the minimal temperature achievable
in a cooling cycle. We have considered a minimal model,
where the purpose is to cool a single qubit and the working
medium is a single mode of the electromagnetic field,
undergoing a (quantum) Otto cycle. 
Since we are interested in the minimum attainable 
temperature, we have considered the most favorable instance 
in which the field is reset to its vacuum state
at the beginning of each cycle. 

Our model can be also interpreted as a 
\emph{collision model}~\cite{gisin2002,gisin2002b,buzek2005,buzek2008,palma2009,palma2012,palma2013} of irreversible
quantum dynamics. Such kind of models were used in 
the literature to analyze the process 
of thermalization of a system in contact with a bath.
In our case, the qubit undergoes a sequence of identical 
collisions, described by the unitary evolution operator $U$ 
of Eq.~(\ref{eq:rhoqn}), with the bath composed of an 
arbitrarily large number of oscillators initially in their ground state.
Our results show that, due to the DCE, the qubit thermalizes
to a nonzero temperature, different from the initial $T=0$ temperature
of the bath of oscillators. 

We emphasize that our results, though obtained for a very simple 
model, are exact, in that the coupled matter-field equations
are numerically integrated, without (i) the RWA approximation and
(ii) the master equation approximation often used in the literature. 
Since the DCE is a generic feature of finite-time quantum electrodynamics
and therefore also of finite-time quantum thermodynamics,
we conjecture that the unattainability of the zero temperature limit,
even in the limit of infinite number of cycles, would remain valid
for any \emph{cyclic quantum} cooling machine. At any rate, our 
results call for a deeper understanding of the relevance of 
the dynamical Casimir effect in quantum thermodynamics. 


\begin{acknowledgments}
G.B. acknowledges the support by MIUR-PRIN project
``Collective quantum phenomena: From strongly correlated systems to
quantum simulators''.
\end{acknowledgments}




\begin{thebibliography}{0}
\bibitem{footnote}
We will consider a model without moving boundaries,
but with a time-dependent matter-field coupling 
$g(t)$. Since the coupling constant $g\propto 1/\sqrt{V}$, with 
$V$ the quantization volume for the field, 
changing $g$ in time is equivalent to changing the volume $V$,
see Ref.~\cite{casimirepjd}.
\bibitem{footnote2}
More precisely, from numerical data as well as from a
second order (in the interaction Hamiltonian $H_I$)
perturbative treatment of finite-time quantum electrodynamics,
outlined in Ref.~\cite{casimirepjd}, we find that
$$
z_\infty=\frac{1}{6}\,(\omega\tau)^2\,\left[1+
\frac{9g^2}{4\omega^2}\right].
$$
\end{thebibliography}
\end{document}